\documentclass[12pt]{article}
\usepackage{amscd,amsfonts,amssymb,amsmath,latexsym,array,hhline}

\renewcommand{\bar}{\overline}
\renewcommand{\tilde}{\widetilde}
\renewcommand{\hat}{\widehat}
\renewcommand{\kappa}{\varkappa}

\newcommand{\LL}{{\mathbb L}}

\newcommand{\NN}{{\mathbb N}}

\numberwithin{equation}{section}

\renewcommand{\section}[1]{\refstepcounter{section}
                            {\noindent\large\bf\thesection. #1}}

\makeatletter

\renewcommand{\section}{\@startsection{section}{1}{0pt}{30pt}{6pt}{\large\bf}}
\def\dot{\hspace{-16pt}.\hspace{-2pt} }

\renewcommand{\@makefnmark}{}
\renewcommand{\@cite}[2]{[{#1\if@tempswa ; #2\fi}]}

\makeatother

\begin{document}

\title{\bf Two switching multiple disorder \\
problems for Brownian motions} 

\author{Pavel V. Gapeev\footnote{London School of Economics, Department of
Mathematics, Houghton Street, London WC2A 2AE, United Kingdom;
e-mail: p.v.gapeev{\char'100}lse.ac.uk}}
\date{}
\maketitle


\begin{abstract}
       The multiple disorder problem seeks to determine a sequence
       of stopping times which are as close as possible to the unknown
       times of disorders at which the observation process changes
       its probability characteristics. We derive closed form solutions
       in two formulations of the multiple disorder problem for an
       observable Brownian motion with switching constant drift rates.
       The method of proof is based on the reduction of the initial problems
       to appropriate optimal switching problems and the analysis of the associated
       coupled free-boundary problems. We also describe the sequential switching multiple
       disorder detection procedures resulting from these formulations.

\end{abstract}






\section{\dot Introduction}

     Suppose that at time $t = 0$ we begin to observe a sample path of some
     continuous process $X=(X_t)_{t \ge 0}$ with probability characteristics
     changing at some unknown disorder times $(\eta_n)_{n \in \NN}$ at which
     an unobservable two-state process $\Theta=(\Theta_t)_{t \ge 0}$ 
     switches between one state and the other. The switching multiple disorder problem
     is to decide at which time instants $(\tau_n)_{n \in \NN}$ one should give alarm
     signals to indicate the occurrence of changes in the current state of the process
     $\Theta$ as close as possible to the initial disorder times $(\eta_n)_{n \in \NN}$.
Such disorder (or change-point) detection problems
have originally arisen and still play a prominent role in quality
control, where one observes the output of a production line and
wishes to detect deviations from acceptable levels. After the
introduction of the original control charts by Shewhart \cite{She},
various modifications of the disorder problem have been recognized
(see, e.g. Pages \cite{Pa}) and implemented in a number of applied
sciences (see, e.g. Carlstein, M\"uller and Siegmund \cite{CMS}).

    The problem of detecting a single change in the constant drift rate of a Brownian motion (Wiener
    process) was formulated and explicitly solved by Shiryaev \cite{S61b}-\cite{S63} and \cite{S65}-\cite{SCyb}
    (see also Shiryaev \cite[Chapter~IV]{S} and Peskir and Shiryaev \cite[Chapter~VI, Section~22]{PSbook} for further references).
    The optimal time of alarm was sought as a stopping time minimising a linear combination of the
    false alarm probability and the average time delay in detecting of the disorder correctly.
    Shiryaev \cite{S61b} and \cite{S63I} also proposed another formulation of the problem
    in which the occurrence of a single change should be preceded by a long period of observations
    under which a stationary regime has been established. The resulting optimal {\it multistage}
    detection procedure consisted in searching for a sequence of stopping times minimising
    the average time delay given that the mean time between two false alarms is fixed.
    More recently, Feinberg and Shiryaev \cite{FeinSh} derived an explicit solution of the quickest
    detection problem in the generalized Bayesian formulation and proved the asymptotic optimality
    of the associated detection procedure for the related minimax formulation.
    Extensive overviews of these and other related sequential quickest change-point
    detection methods were provided in Shiryaev \cite{S02} and Poor and Hadjiliadis \cite{PH}. 

     In the present paper, we formulate and solve the switching multiple disorder problem for an observed
     Wiener process $X$ changing its drift rate from $\mu_j$ to $\mu_{1-j}$ when $\Theta$
     changes its state from $j$ to $1-j$, for every $j = 0, 1$. In contrast to the problem of detecting a {\it single}
     change, in the {\it switching multiple} disorder problem, one looks for an infinite sequence of the alarm times
     $(\tau_n)_{n \in \NN}$ minimising a series of linear combinations of {\it discounted} average losses due to false
     alarms and delay penalties in detecting of the disorder times $(\eta_n)_{n \in \NN}$ correctly.
     We propose two different formulations of the problem based on a specification of dynamics of the process $\Theta$.
     In the first formulation, $\Theta$ is assumed to be a continuous time Markov chain of intensity $\lambda$,
     the dynamics of which are not influenced by the alarm times $(\tau_n)_{n \in \NN}$.
     In the second formulation, it is assumed that the subsequent time $\eta_n$, at which $\Theta$ changes its state,
     can only occur after the previous alarm is sounded at $\tau_{n-1}$. Moreover, it is assumed that the differences
     $(\eta_n - \tau_{n-1})_{n \in \NN}$ form a sequence of (conditionally) independent exponential random variables.

Apart from other possible areas of application, such a situation
usually happens in models of liquid financial markets having trading
investors of different kinds.
It is natural to assume that the {\it small} investors can only
influence little fluctuations of the market prices of risky assets,
while the {\it large} investors can affect the pricing trends as
well, by means of either buying or selling substantial amounts of
assets. More precisely, the pricing trends should either rise up or
fall down at some random times, after essential amounts of assets
are bought or sold, respectively.
We can thus consider a model of a financial market of such kind
in which the dynamics (of the logarithms) of the asset prices are
described by a Brownian motion with switching drift rates. We may
further assume that our model allows for an infinite number of
transactions (free on charge) on the infinite time interval and use
an exponential constant discounting rate $r$, which can be chosen
equal to the riskless short rate of a bank account. The problem of
detecting of a single change in the probability characteristics of
accessible financial data, which is associated with the appearance
of arbitrage opportunities in the market, was considered by Shiryaev
\cite{S02}.

In the present paper, we reduce the initial multiple disorder
problems to appropriate {\it optimal switching problems} for
filtering estimates of the current state of the unobservable drift
rate of a Brownian motion. The use of exponential discounting makes
our problem well connected to the problem of single disorder
detection with exponential delay penalty costs studied by Poor
\cite{Poor}, Beibel \cite{Beibel}, and Bayraktar and Dayanik
\cite{BD}.
We show that the optimal switching times can be expressed as the
first times at which the appropriate posterior probability processes
hit certain constant boundaries. We derive closed form expressions
for the resulting Bayesian risk functions and the optimal switching
boundaries by means of solving the associated {\it coupled
free-boundary problems} for ordinary differential operators. We also
construct sequential switching multiple disorder detection
procedures resulting from the two formulations.

     Optimal switching problems represent extensions of stopping problems and
     games in which one looks for an infinite sequence of optimal stopping times.
     A general approach for studying such problems was developed
     in Bensoussan and Friedman \cite{BenFr1}-\cite{BenFr2} and
     Friedman \cite{Fr3} (see also Friedman \cite[Chapter~XVI]{Fr}).
     This investigation was continued by
     Brekke and {\O}ksendal \cite{BO}, Duckworth and Zervos
     \cite{DZ}, Yushkevich and Gordienko \cite{YG},
     and Hamad{\`e}ne and Jeanblanc \cite{HJ} among others for the continuous time case,
     and by Yushkevich \cite{Y1}-\cite{Y2} for the discrete time case.
Other optimal switching and impulse control problems involving
hidden Markov chains in the observable jump processes were recently
studied by Bayraktar and Ludkovski
\cite{BayLudk09}-\cite{BayLudk10}.


       The paper is organized as follows. In Section 2, for the initial multiple
       disorder problems, we construct the appropriate optimal switching problems and reduce
       the latter to their equivalent coupled optimal stopping problems.
       In Section 3, we derive closed form solutions of the associated coupled
       free-boundary problems, which are expressed in terms of Heun's double
       confluent functions and Kummer's confluent hypergeometric functions.
       In Section 4, we verify that the solutions of the coupled free-boundary problems
       provide the solutions of the initial optimal switching problems, and
       describe the resulting sequential switching multiple disorder detection procedures.
       The main results of the paper are stated in Theorems 4.1 and 4.2.
       The optimal sequential detecting schemes are displayed more explicitly in Remark 4.3.

  \section{\dot Formulation of the problems}


  In this section, we give two formulations of
  the switching multiple disorder problem for an observed Brownian motion
  (see, e.g. \cite[Chapter~IV, Section~4]{S} or \cite[Chapter~VI, Section~22]{PSbook} for the single disorder case).
  In these formulations, it is assumed that one observes a sample path of the Brownian motion $X$ with the drift
  rate switching between $\mu_0$ and $\mu_1$ at some random times $(\eta_n)_{n \in \NN}$.

  \vspace{6pt}


  2.1. (The setting.)
  Let us assume that all the considerations take place on a probability
  space $(\Omega, {\cal G}, {P}_{\pi})$ with a
  standard Brownian motion (Wiener process) $B=(B_t)_{t \ge 0}$ started at zero under $P_{\pi}$.
  Suppose that there exists a right-continuous process $\Theta$ with two states
  $0$ and $1$, having the initial distribution $\{1-\pi, \pi \}$ under $P_{\pi}$, for $\pi \in [0, 1]$.
  It is assumed that the process $\Theta$ is unobservable, so that
  the switching times $\eta_n = \inf \{ t \ge \eta_{n-1} \, | \,
  \Theta_t \neq \Theta_{\eta_{n-1}} \}$, for $n \in \NN$, with $\eta_0 = 0$,
  at which $\Theta$ changes its state from $j$ to $1-j$, for every $j = 0, 1$,
  are unknown, that is, they cannot be observed directly.

  Suppose that we observe a continuous process $X=(X_t)_{t \ge 0}$
  solving the stochastic differential equation:
  \begin{equation}
  \label{X4}
  dX_t = \big( \mu_0 + (\mu_1 - \mu_0) \, \Theta_t \big) \, dt + \sigma \, dB_t \quad (X_0 = 0)
  \end{equation}
  where $\mu_0 \neq \mu_1$ and $\sigma > 0$ are some given constants.
  Being based upon the continuous observation of $X$,
  our task is to find among non-decreasing sequences of stopping times
  $(\tau_n)_{n \in \NN}$ of $X$ (i.e., stopping times with respect to
  the natural filtration ${\cal F}_t=\sigma (X_s \, | \, 0 \le s \le t)$
  of the process $X$, for $t \ge 0$) at which the alarms should be sounded {\it as close as possible}
  to the unknown switching times of the process $\Theta$. 
  More precisely, the {\it switching multiple disorder problem} consists of computing
  the Bayesian risk functions:
  \begin{align}
  \label{V0}
  V_0^*(\pi)=\inf_{(\tau_{0,n})} \sum_{k = 1}^{\infty} E_{\pi} \bigg[
  &a \, e^{- r \tau_{0, 2k-1}} \, I(\Theta_{\tau_{0, 2k-1}} = 1) +
  b \, e^{- r \tau_{0, 2k}} \, I(\Theta_{\tau_{0, 2k}} = 0) \\
  \notag
  &+ \sum_{j=0}^1 \int_{\tau_{0, 2k-2+j}}^{\tau_{0, 2k-1+j}} e^{- r t} \, I(\Theta_t = j) \, dt \bigg] \\
  \label{W0}
  V_1^*(\pi)=\inf_{(\tau_{1,n})} \sum_{k = 1}^{\infty} E_{\pi} \bigg[
  &b \, e^{- r \tau_{1, 2k-1}} \, I(\Theta_{\tau_{1, 2k-1}} = 0) +
  a \, e^{- r \tau_{1, 2k}} \, I(\Theta_{\tau_{1, 2k}} = 1) \\
  \notag
  &+ \sum_{j=0}^1 \int_{\tau_{1, 2k-2+j}}^{\tau_{1, 2k-1+j}} e^{- r t} \, I(\Theta_t = 1-j) \, dt \bigg]
  \end{align}
  and finding the non-decreasing sequences of optimal
  stopping times $(\tau^*_{i, n})_{n \in \NN}$ such that $\tau^*_{i, 0} = 0$, $i = 0, 1$, 
  at which the infima in (\ref{V0}) and (\ref{W0}) are attained,
  respectively, where $I(\cdot)$ denotes the indicator function.
  Note that the function $V_i^*(\pi)$ expresses the Bayesian risk of the whole sequence
  $(\tau_{i,n})_{n \in \NN}$ in the case in which the process $\Theta$
  starts at $\Theta_0 = 1-i$, for every $i = 0, 1$ fixed, and all $\pi \in [0, 1]$.
  We therefore see that $E_{\pi} \big[e^{- r \tau_{i,n}} \, I(\Theta_{\tau_{i,n}} = j) \big]$
  is the average discounted loss due to a false alarm, and
  $E_{\pi} \big[ \int_{\tau_{i,n-1}}^{\tau_{i,n}} e^{- r t} \, I(\Theta_t=1-j) \, dt \big]$
  expresses the average discounted loss due to a delay in detecting of the time at which $\Theta$
  changes its state from $j$ to $1-j$ correctly, for every $i, j = 0, 1$ and any $n \in \NN$.
  In this case, $a, b > 0$ are costs of false alarms and $r > 0$ is a discounting rate.


  Using the fact that $(\tau_{i,n})_{n \in \NN}$ is a non-decreasing sequence of stopping times
  with respect to the filtration $({\cal F}_t)_{t \ge 0}$, by means of standard arguments, which
  are similar to those presented in \cite[pages~195-197]{S}, we get that:
  \begin{equation}
  \label{J4}
  E_{\pi} \big[ e^{- r \tau_{i,n}} \, I(\Theta_{\tau_{i,n}} = j) \big] =
  E_{\pi} \big[ E_{\pi} \big[ e^{- r \tau_{i,n}} \, I(\Theta_{\tau_{i,n}} = j) \, \big| \, {\cal F}_{\tau_{i,n}} \big] \big] =
  E_{\pi} \big[ e^{- r \tau_{i,n}} \, P_{\pi}(\Theta_{\tau_{i,n}} = j \, | \, {\cal F}_{\tau_{i,n}}) \big]
  \end{equation}
  and
  \begin{align}
  \label{I4}
  &E_{\pi} \bigg[ \int_{\tau_{i,n-1}}^{\tau_{i,n}} e^{- r t} \, I(\Theta_{\tau_{i,n}} = j) \, dt \bigg]
  = E_{\pi} \bigg[ \int_0^{\infty} e^{- r t} \, I(\tau_{i,n-1} \le t, \Theta_t = j, t < \tau_{i,n}) \, dt \bigg] \\
  \notag
  &= E_{\pi} \bigg[ \int_0^{\infty} E_{\pi} \big[ e^{- r t} \,  I(\tau_{i,n-1} \le t, \Theta_t = j, t < \tau_{i,n}) \, \big|
  \, {\cal F}_t \big] \, dt \bigg]
  = E_{\pi} \bigg[ \int_{\tau_{i,n-1}}^{\tau_{i,n}} e^{- r t} \, P_{\pi} (\Theta_t = j \, | \, {\cal F}_t) \, dt \bigg]
  \end{align}
  holds for every $i, j = 0, 1$ and any $n \in \NN$.

We further consider two different formulations of the problem, depending on the specified dynamics of the process $\Theta$. The first
formulation does not involve any influence of the alarm times
$\tau_{i,n}$ on the times of changes $\eta_{n}$. In the second
formulation, it is assumed that the change at $\eta_{n}$ can occur
only after the previous alarm is sounded at $\tau_{i,n-1}$, for
every $i = 0, 1$ and any $n \in \NN$.


\vspace{6pt}

  2.2. (The first formulation.)
  Suppose that $\Theta$ is a continuous time Markov chain which is independent of the Brownian motion $B$
  and has the initial distribution $\{ 1-\pi, \pi \}$ under $P_{\pi}$. Assume that $\Theta$ has the
  transition-probability matrix $\{ e^{-\lambda t}, 1- e^{-\lambda t}; 1 - e^{-\lambda t}, e^{-\lambda t} \}$
  and the intensity-matrix $\{ -\lambda, \lambda; \lambda, -\lambda \}$, for all $t \ge 0$ and some $\lambda > 0$ fixed.
  In other words, the Markov chain $\Theta$ changes its state at exponentially distributed times of intensity $\lambda$,
  which are independent of the dynamics of the Brownian motion $B$.
  Such a process $\Theta$ is called {\it telegraphic signal} of intensity $\lambda$
  in the literature (see, e.g. \cite[Chapter~IX, Section~4]{LS} or \cite[Chapter~VIII]{EAM}).

  It thus follows from \cite[Chapter~IX, Theorem~9.1]{LS} (see also \cite[Chapter~IX, Example~3]{LS})
  that the {\it posterior probability} process $\Pi = (\Pi_t)_{t \ge 0}$
  defined by $\Pi_t = P_{\pi} (\Theta_t = 1 \, | \, {\cal F}_t)$ solves the stochastic differential equation:
  \begin{equation}
  \label{pi3c}
  d\Pi_t = \lambda (1-2\Pi_t) \, dt + \frac{\mu_1 - \mu_0}{\sigma} \, \Pi_t(1-\Pi_t) \, d{\bar B}_t \quad (\Pi_0 = \pi)
  \end{equation}
  where the innovation process ${\bar B} = ({\bar B}_t)_{t \ge 0}$ defined by:
  \begin{equation}
  \label{barW3}
  {\bar B}_t = \frac{1}{\sigma} \bigg(X_t - \int_0^t \big( \mu_{0} + (\mu_{1}-\mu_{0}) \, \Pi_s \big) \, ds \bigg)
  \end{equation}
  is a standard Brownian motion according to P. L{\'e}vy's characterization theorem (see, e.g. \cite[Chapter~IV, Theorem~4.1]{LS}).
  It is also seen from (\ref{pi3c}) that $\Pi$ is a (time-homogeneous strong) Markov process with respect to
  its natural filtration, which obviously coincides with $({\cal F}_t)_{t \ge 0}$.


  Taking into account the expressions in (\ref{J4}) and (\ref{I4}), we therefore conclude
  that the Bayesian risk functions from (\ref{V0}) and (\ref{W0}) admit the representations:
  \begin{align}
  \label{V4a}
  V_0^*(\pi) = \inf_{(\tau_{0,n})} \sum_{k = 1}^{\infty}
  E_{\pi} \bigg[ &a \, e^{- r \tau_{0,2k-1}} \, \Pi_{\tau_{0,2k-1}}
  + \int_{\tau_{0,2k-2}}^{\tau_{0,2k-1}} e^{- r t} \, (1 - \Pi_t) \, dt \\
  \notag
  &+ b \, e^{- r \tau_{0,2k}} \, (1-\Pi_{\tau_{0,2k}})
  + \int_{\tau_{0,2k-1}}^{\tau_{0,2k}} e^{- r t} \, \Pi_t \, dt \bigg] \\
  \label{W4a}
  V_1^*(\pi) = \inf_{(\tau_{1,n})} \sum_{k = 1}^{\infty}
  E_{\pi} \bigg[ &b \, e^{- r \tau_{1,2k-1}} \, (1-\Pi_{\tau_{1,2k-1}})
  + \int_{\tau_{1,2k-2}}^{\tau_{1,2k-1}} e^{- r t} \, \Pi_t \, dt \\
  \notag
  &+ a \, e^{- r \tau_{1,2k}} \, \Pi_{\tau_{1,2k}}
  + \int_{\tau_{1,2k-1}}^{\tau_{1,2k}} e^{- r t} \, (1 - \Pi_t) \, dt \bigg]
  \end{align}
  where the infima are taken over all sequences of stopping times
  $(\tau_{i,n})_{n \in \NN}$, $i = 0, 1$, of the process $\Pi$.
  %
  By virtue of the strong Markov property of the process $\Pi$,
  we can reduce the system of (\ref{V4a}) and (\ref{W4a})
  to the following {\it coupled optimal stopping problem}:
  \begin{align}
  \label{V4b}
  V_0^*(\pi) &= \inf_{\tau_0} E_{\pi} \left[
  a \, e^{- r \tau_0} \, \Pi_{\tau_0} + \int_0^{\tau_0} e^{- r t} \, (1-\Pi_t) \, dt + V_1^*(\Pi_{\tau_0}) \right]  \\
  \label{W4b}
  V_1^*(\pi) &= \inf_{\tau_1} E_{\pi} \left[
  b \, e^{- r \tau_1} \, (1-\Pi_{\tau_1}) + \int_0^{\tau_1} e^{- r t} \, \Pi_t \, dt + V_0^*(\Pi_{\tau_1}) \right]
  \end{align}
  where the infima 
  are taken over all stopping times $\tau_i$, $i = 0, 1$, of the process $\Pi$ with $P_{\pi}(\Pi_0 = \pi)=1$.
  We further search for optimal stopping times
  in (\ref{V4b}) and (\ref{W4b}) of the form:
  \begin{equation}
  \label{tau13a}
  \tau^*_0 = \inf \{ t \ge 0 \, | \, \Pi_t \le g_* \} \quad
  \text{and} \quad \tau^*_1 = \inf \{ t \ge 0 \, | \, \Pi_t \ge h_* \}
  \end{equation}
  for some $0 < g_* < h_* < 1$,
  where $g_*$ is the largest and $h_*$ is the smallest number $\pi$ from $[0, 1]$
  such that $V_0^*(\pi) = a \pi + V_1^*(\pi)$ and $V_1^*(\pi) = b (1-\pi) + V_0^*(\pi)$
  holds, respectively. This fact implies that the sequences of stopping times
  $(\tau^*_{i,n})_{n \in \NN}$ given by:
  \begin{equation}
  \label{tau13}
  \tau^*_{i,2k-1+i} = \inf \{t \ge \tau^*_{i,2k-2+i} \, | \, \Pi_t \le g_* \}
  \quad \text{and} \quad
  \tau^*_{i,2k-i} = \inf \{t \ge \tau^*_{i,2k-1-i} \, | \, \Pi_t \ge h_* \}
  \end{equation}
  for every $i = 0, 1$ and any $k \in \NN$, are optimal in the problems of (\ref{V4a}) and (\ref{W4a}).

\vspace{6pt}

  2.3. (The second formulation.) As that is the case in the previous formulation, for every $i = 0, 1$, let us denote by $(\zeta_{i, 2k-i})_{k \in
  \NN}$ and $(\zeta_{i, 2k-1+i})_{k \in \NN}$ the sequences of alarm times sounded to indicate that the state of
  $\Theta$ has been changed from $0$ to $1$ or from $1$ to $0$, respectively. 
  Let us now assume that the switching time $\eta_{n}$ of the process $\Theta$ can only occur after the previous
  alarm is sounded at $\zeta_{i,n-1}$, for any $n \in \NN$. 
  Suppose that $(\xi_{i,n})_{n \in \NN}$ defined by $\xi_{i,n} = \eta_n - \tau_{i,n-1}$ forms a sequence of (conditionally) independent
  non-negative random variables such that $\xi_{i,n}$ is independent of the Brownian motion $B$ on the time interval $[\zeta_{i,n-1}, \zeta_{i,n}]$.
  Moreover, we assume that the properties $P_{\pi}(\eta_n = \zeta_{i,n-1} \, | \, {\cal F}_{\zeta_{i,n-1}}) =
  \Pi_{\zeta_{i,n-1}}$ and $P_{\pi}(\eta_n > t \, | \, \eta_n > \zeta_{i,n-1},
  {\cal F}_{\zeta_{i,n-1}}) = e^{- \lambda (t - \zeta_{i,n-1})}$ hold for all
  $t \ge \zeta_{i,n-1}$ and some $\lambda > 0$ fixed.
  In other words, the process $\Theta$ changes its state in the exponential time $\xi_{i,n} = \eta_n - \zeta_{i,n-1}$ of intensity $\lambda$
  after the time of the previous alarm $\zeta_{i,n-1}$, where $\xi_{i,n}$
  does not depend on the subsequent fluctuations of the process $B$.

  It thus follows from \cite[Chapter~IX, Theorem~9.1]{LS} (see also \cite[Chapter~IX, Example~2]{LS} or \cite[Chapter~VIII]{EAM})
  that the posterior probability process $\Pi$ solves the stochastic differential equation:
  \begin{equation}
  \label{pi4ca}
  d\Pi^{(0)}_t = - \lambda \Pi^{(0)}_t \, dt + \frac{\mu_1 - \mu_0}{\sigma} \,
  \Pi^{(0)}_t(1-\Pi^{(0)}_t) \, d{\bar B}_t \quad (\Pi^{(0)}_{\zeta_{i,2k-2+i}} = \Pi^{(1)}_{\zeta_{i,2k-2+i}})
  \end{equation}
  for $\zeta_{i,2k-2+i} \le t \le \zeta_{i,2k-1+i}$ and
  \begin{equation}
  \label{pi4cb}
  d\Pi^{(1)}_t = \lambda (1-\Pi^{(1)}_t) \, dt + \frac{\mu_1 - \mu_0}{\sigma} \,
  \Pi^{(1)}_t(1-\Pi^{(1)}_t) \, d{\bar B}_t \quad (\Pi^{(1)}_{\zeta_{i,2k-1-i}} = \Pi^{(0)}_{\zeta_{i,2k-1-i}})
  \end{equation}
  for $\zeta_{i,2k-1-i} \le t \le \zeta_{i,2k-i}$, where the process ${\bar B}$
  is defined in (\ref{barW3}) and turns out to be a standard Brownian motion on the time intervals
  $[\zeta_{i,n-1}, \zeta_{i,n}]$, for every $i = 0, 1$ and any $k, n \in \NN$.


  Taking into account the expressions in (\ref{J4}) and (\ref{I4}),
  we may conclude that the Bayesian risk functions in this
  formulation are given by:
  \begin{align}
  \label{V3a}
  U_0^*(\pi) = \inf_{(\zeta_{0,n})} \sum_{k = 1}^{\infty}
  E_{\pi} \bigg[ &a \, e^{- r \zeta_{0,2k-1}} \, \Pi^{(0)}_{\zeta_{0,2k-1}}
  + \int_{\zeta_{0,2k-2}}^{\zeta_{0,2k-1}} e^{- r t} \, (1 - \Pi^{(0)}_t) \, dt \\
  \notag
  &+ b \, e^{- r \zeta_{0,2k}} \, (1-\Pi^{(1)}_{\zeta_{0,2k}})
  + \int_{\zeta_{0,2k-1}}^{\zeta_{0,2k}} e^{- r t} \, \Pi^{(1)}_t \, dt \bigg] \\
  \label{W3a}
  U_1^*(\pi) = \inf_{(\zeta_{1,n})} \sum_{k = 1}^{\infty}
  E_{\pi} \bigg[ &b \, e^{- r \zeta_{1,2k-1}} \, (1-\Pi^{(1)}_{\zeta_{1,2k-1}})
  + \int_{\zeta_{1,2k-2}}^{\zeta_{1,2k-1}} e^{- r t} \, \Pi^{(1)}_t \, dt \\
  \notag
  &+ a \, e^{- r \tau_{1,2k}} \, \Pi^{(0)}_{\zeta_{1,2k}}
  + \int_{\zeta_{1,2k-1}}^{\zeta_{1,2k}} e^{- r t} \, (1 - \Pi^{(0)}_t) \, dt \bigg]
  \end{align}
  where the infima are taken over all sequences of stopping times $(\zeta_{i,n})_{n \in \NN}$
  of the processes $\Pi^{(i)} = (\Pi^{(i)}_t)_{t \ge 0}$, $i = 0, 1$, solving the stochastic
  differential equations in (\ref{pi4ca}) and (\ref{pi4cb}), respectively.
  %
  By virtue of the strong Markov property of the processes $\Pi^{(i)}$, $i = 0, 1$,
  we can reduce the system of (\ref{V3a}) and (\ref{W3a}) to the following
  coupled optimal stopping problem:
  \begin{align}
  \label{V3b}
  U_0^*(\pi) &= \inf_{\zeta_0} E_{\pi} \left[
  a \, e^{- r \zeta_0} \, \Pi^{(0)}_{\zeta_0} + \int_0^{\zeta_0} e^{- r t} \, (1-\Pi^{(0)}_t) \, dt + U_1^*(\Pi^{(0)}_{\zeta_0}) \right]  \\
  \label{W3b}
  U_1^*(\pi) &= \inf_{\zeta_1} E_{\pi} \left[
  b \, e^{- r \zeta_1} \, (1-\Pi^{(1)}_{\zeta_1}) + \int_0^{\zeta_1} e^{- r t} \, \Pi^{(1)}_t \, dt + U_0^*(\Pi^{(1)}_{\zeta_1}) \right]
  \end{align}
  where the infima 
  are taken over all stopping times $\zeta_i$ of the processes $\Pi^{(i)}$, $i = 0, 1$, respectively.
  We further search for the optimal stopping times
  in (\ref{V3b}) and (\ref{W3b}) of the form:
  \begin{equation}
  \label{tau13a4}
  \zeta^*_0 = \inf \{ t \ge 0 \, | \, \Pi^{(0)}_t \le p_* \}
  \quad \text{and} \quad
  \zeta^*_1 = \inf \{ t \ge 0 \, | \, \Pi^{(1)}_t \ge q_* \}
  \end{equation}
  for some $0 < p_* < q_* < 1$,
  where $p_*$ is the largest and $q_*$ is the smallest number $\pi$ from $[0, 1]$
  such that $U_0^*(\pi) = a \pi + U_1^*(\pi)$ and $U_1^*(\pi) = b (1-\pi) + U_0^*(\pi)$
  holds, respectively. This fact implies that the sequences of stopping times
  $(\zeta^*_{i,n})_{n \in \NN}$ given by:
  \begin{equation}
  \label{sigma13}
  \zeta^*_{i,2k-1+i} = \inf \{t \ge \zeta^*_{i,2k-2+i} \, | \, \Pi^{(0)}_t \le p_* \}
  \quad \text{and} \quad
  \zeta^*_{i,2k-i} = \inf \{t \ge \zeta^*_{i,2k-1-i} \, | \, \Pi^{(1)}_t \ge q_* \}
  \end{equation}
  for every $i = 0, 1$ and any $k \in \NN$, are optimal in the problems of (\ref{V3a}) and (\ref{W3a}).

\vspace{6pt}

  {\bf Remark 2.1.} Recall that $(\zeta_{i,n})_{n \in \NN}$ is a
  non-decreasing sequence of stopping times with respect to the
  filtration $({\cal F}_t)_{t \ge 0}$. Then, by virtue of the
  assumption that $\eta_n \ge \zeta_{i, n-1}$ holds,
  standard arguments show that the equalities: 
  \begin{align}
  \label{I5}
  &E_{\pi} \bigg[ \int_{\zeta_{i,n-1}}^{\zeta_{i,n}} e^{- r t} \, I(\Theta_t = j) \, dt \bigg]
  =E_{\pi} \bigg[ \int_{0}^{\infty} e^{- r t} \, I(\zeta_{i,n-1} \le t, \eta_n \le t, t < \zeta_{i,n}) \, dt \bigg]
  \\ \notag
  &=E_{\pi} \bigg[ I({\eta_n} < {\zeta_{i,n}}) \int_{\eta_n}^{\zeta_{i,n}} e^{- r t} \, dt \bigg]
  = \frac{1}{r} \, E_{\pi} \Big[  e^{- r \zeta_{i,n}} \, \big(e^{r (\zeta_{i,n} - \eta_n)^+} - 1 \big) \Big]
  \end{align}
  are satisfied for every $i, j = 0, 1$ and any $n \in \NN$.
  This fact builds a connection between the introduction of exponential
  discounting into the switching multiple disorder problem of (\ref{V3a})-(\ref{W3a})
  and the consideration of single disorder detection problems with exponential delay
  penalty costs studied in \cite{Poor}, \cite{Beibel} and \cite{BD}.

       \vspace{6pt}

        2.4. (Coupled free-boundary problems.)
        Standard arguments based on an application of It{\^o}'s formula
        (see, e.g. \cite[Chapter~V, Section~5.1]{KS} or \cite[Chapter~VII, Section~7.3]{O})
        imply that the infinitesimal operator $\LL$ of the process $\Pi$ 
        from (\ref{pi3c}) acts on an arbitrary twice continuously differentiable (locally) bounded function 
        $F(\pi)$ according to the rule:
        \begin{equation}
        \label{Lf5}
        (\LL F)(\pi) = \lambda \, (1 - 2 \pi) \, F'(\pi) +
        \frac{1}{2} \, \Big( \frac{\mu_1-\mu_0}{\sigma} \Big)^2 \, \pi^2 (1-\pi)^2 \, F''(\pi)
        \end{equation}
        for all $\pi \in (0, 1)$. In order to find the unknown value functions
        $V_0^*(\pi)$ and $V_1^*(\pi)$ from (\ref{V3b}) and (\ref{W3b})
        as well as the unknown boundaries $g_*$ and $h_*$
        from (\ref{tau13a}), 
        we may use the results of the general theory of optimal stopping problems for
        continuous time Markov processes (see, e.g. 
        \cite[Chapter~III, Section~8]{S} and \cite[Chapter~IV, Section~8]{PSbook}).
        More precisely, we formulate the associated {\it coupled free-boundary problem}:
    \begin{align}
    \label{VC3}
    (\LL V_0 - r V_0)(\pi) = - (1-\pi) \;\;\; \text{for} \;\;\; \pi > g,
    \quad &(\LL V_1 - r V_1)(\pi) = - \pi \;\;\; \text{for} \;\;\; \pi < h \\
    \label{instop3}
    V_0(g+) = a \, g + V_1(g+), \quad &V_1(h-) = b \, (1-h) + V_0(h-) \\
    \label{smfit3}
    V_0'(g+) = a + V_1'(g+), \quad &V_1'(h-) = - b + V_0'(h-) \\
    \label{Upi3a}
    V_0(\pi) = a \, \pi + V_1(\pi)  \;\;\; \text{for} \;\;\; \pi < g, \quad
    &V_1(\pi) = b \, (1-\pi) + V_0(\pi) \;\;\; \text{for} \;\;\; \pi > h \\
    \label{Cont3c}
    V_0(\pi) < a \, \pi + V_1(\pi) \;\;\; \text{for} \;\;\; \pi > g, \quad
    &V_1(\pi) < b \, (1-\pi) + V_0(\pi) \;\;\; \text{for} \;\;\; \pi < h \\
    \label{VD3}
    (\LL V_0 - r V_0)(\pi) > - (1-\pi) \;\;\; \text{for}
    \;\;\; \pi < g, \quad &(\LL V_1 - r V_1)(\pi) > - \pi \;\;\;
    \text{for} \;\;\; \pi > h
    \end{align}
    with $0 < g < h < 1$,
    where the {\it instantaneous-stopping}
    and {\it smooth-fit} conditions of (\ref{instop3})
    and (\ref{smfit3}) are satisfied at $g_*$ and $h_*$.
    Note that the superharmonic characterisation of the value function
    (see, e.g. \cite[Chapter~III, Section~8]{S} and \cite[Chapter~IV, Section~9]{PSbook})
    implies that $V_0^*(\pi)$ from (\ref{V4b}) and $V_1^*(\pi)$ from (\ref{W4b})
    are the largest functions satisfying the expressions in (\ref{VC3})-(\ref{instop3}) and
    (\ref{Upi3a})-(\ref{Cont3c}) with the boundaries $g_*$ and $h_*$.

        Furthermore, standard arguments show that the infinitesimal
        operator $\LL_i$ of the process $\Pi^{(i)}$ from (\ref{pi4ca})-(\ref{pi4cb})
        acts on an arbitrary twice continuously differentiable (locally) bounded function 
        $F(\pi)$ according to the rule:
        \begin{align}
        \label{Lf3}
        (\LL_0 F)(\pi) &= - \lambda \, \pi \, F'(\pi) +
        \frac{1}{2} \, \Big( \frac{\mu_1-\mu_0}{\sigma} \Big)^2 \, \pi^2 (1-\pi)^2 \, F''(\pi) \\
        \label{Lf4}
        (\LL_1 F)(\pi) &= \lambda \, (1 - \pi) \, F'(\pi) +
        \frac{1}{2} \, \Big( \frac{\mu_1-\mu_0}{\sigma} \Big)^2 \, \pi^2 (1-\pi)^2 \, F''(\pi)
        \end{align}
        for all $\pi \in (0, 1)$ and every $i = 0, 1$.
        In order to find the unknown value functions
        $U_0^*(\pi)$ and $U_1^*(\pi)$ from (\ref{V3b}) and (\ref{W3b})
        as well as the unknown boundaries $p_*$ and $q_*$ from (\ref{tau13a}),
        we formulate the associated coupled free-boundary problem:
    \begin{align}
    \label{VC4}
    (\LL_0 U_0 - r U_0)(\pi) = - (1-\pi) \;\;\; \text{for} \;\;\; \pi > p,
    \quad &(\LL_1 U_1 - r U_1)(\pi) = - \pi \;\;\; \text{for} \;\;\; \pi < q \\
    \label{instop4}
    U_0(p+) = a \, p + U_1(p+), \quad &U_1(q-) = b \, (1-q) + U_0(q-) \\
    \label{smfit4}
    U_0'(p+) = a + U_1'(p+) , \quad &U_1'(q-) = - b + U_0'(q-) \\
    \label{Upi4a}
    U_0(\pi) = a \, \pi + U_1(\pi) \;\;\; \text{for} \;\;\; \pi < p, \quad
    &U_1(\pi) = b \, (1-\pi) + U_0(\pi) \;\;\; \text{for} \;\;\; \pi > q \\
    \label{Cont4c}
    U_0(\pi) < a \, \pi + U_1(\pi)  \;\;\; \text{for} \;\;\; \pi > p, \quad
    &U_1(\pi) < b \, (1-\pi) + U_0(\pi) \;\;\; \text{for} \;\;\; \pi < q \\
    \label{VD4}
    (\LL_0 U_0 - r U_0)(\pi) > - (1-\pi) \;\;\; \text{for} \;\;\; \pi < p,
    \quad &(\LL_1 U_1 - r U_1)(\pi) > - \pi \;\;\; \text{for} \;\;\; \pi > q
    \end{align}
    with $0 < p < q < 1$,
    where the {\it instantaneous-stopping} and {\it smooth-fit} conditions of (\ref{instop4})
    and (\ref{smfit4}) are satisfied at $p_*$ and $q_*$.
    The superharmonic characterisation of the value function
    implies that $U_0^*(\pi)$ from (\ref{V3b}) and $U_1^*(\pi)$ from (\ref{W3b})
    are the largest functions satisfying the expressions in (\ref{VC4})-(\ref{instop4}) and (\ref{Upi4a})-(\ref{Cont4c})
    with the boundaries $p_*$ and $q_*$.

 \section{\dot Solutions of the coupled free-boundary problems}

      In this section we solve the systems of (\ref{VC3})-(\ref{VD3})
      and (\ref{VC4})-(\ref{VD4}) and prove the existence and uniqueness
      of solutions of those coupled free-boundary problems associated
      to the corresponding formulations of the switching multiple disorder problem.

      \vspace{6pt}

       3.1. (Existence in the first formulation.)
       The general solutions of the second order ordinary differential equations in (\ref{VC3}) are given by:
       \begin{equation}
       \label{A33c}
       V_i(\pi) = C_{i0} \, Q_{i0}(\pi) + C_{i1} \, Q_{i1}(\pi) + \frac{\lambda}{r(2\lambda + r)}
       + \frac{i \pi}{2\lambda + r} + \frac{(1-i)(1 - \pi)}{2\lambda + r}
       \end{equation}
       where $C_{ij}$, $j = 0, 1$, are some arbitrary constants,
       and the functions $Q_i(\pi)$, $i = 0, 1$, are given by:
        \begin{equation}
        \label{Q01}
        Q_i(\pi) = \sqrt{\pi (1-\pi)} \, \exp \left( \frac{i 2 \lambda}{\rho (1-\pi)}
        + \frac{(1-i) 2 \lambda}{\rho \pi} \right) \, H_i \left((-1)^{i+1} \, \varphi, \psi, 0, \xi; \frac{1}{1-2\pi} \right)
        \end{equation}
        for all $\pi \in (0, 1)$ with
        \begin{equation}
        \label{rho}
        \rho = \left( \frac{\mu_1-\mu_0}{\sigma} \right)^2, \quad \varphi = \frac{8 \lambda}{\rho}, \quad
        \psi = \frac{\varphi^2}{4} + \varphi - \frac{8r}{\rho} - 1 \quad \text{and} \quad
        \xi = 4 \varphi - \psi.
        \end{equation}
        Here, the functions $H_i(\alpha, \beta, \gamma, \delta; x)$, $i = 0, 1$,
        are two positive fundamental solutions (i.e. nontrivial linearly independent particular solutions) of
        Heun's double confluent ordinary differential equation: 
        \begin{equation}
        \label{Heun}
        H''(x) + \frac{2 x^5 - \alpha x^4 - 4 x^3 + 2 x + \alpha}{(x-1)^3(x+1)^3} \, H'(x) +
        \frac{\beta x^2 + (2 \alpha + \gamma) x + \delta}{(x-1)^3(x+1)^3} \, H(x) = 0
        \end{equation}
        with the boundary conditions $H(0) = 1$ and $H'(0) = 0$. Note
        that the series expansion of the solution of the equation in
        (\ref{Heun}) converges under all $-1 < x < 1$,
        and the appropriate analytic continuation can be obtained through the identity
        $H(\alpha, \beta, \gamma, \delta; x) = H(-\alpha, -\delta, -\gamma, -\beta;
        1/x)$. 
        The (irregular) singularities at $-1$ and $1$ of the equation in (\ref{Heun})
        are of unit rank and can be transformed into that of a confluent hypergeometric equation
        (see, e.g. \cite{Decarreau} and \cite{Ronveaux} for an extensive overview and further details).
        According to the results from \cite[Chapter~V, Section~50]{RW}, we can specify the positive
        (strictly) convex functions $Q_i(\pi)$, $i = 0, 1$, as (strictly) decreasing and
        increasing on the interval $(0, 1)$ and having singularities at $0$ and $1$, respectively.

Taking into account the fact that the function $V_0(\pi)$ should be
bounded as $\pi \uparrow 1$ while the function $V_1(\pi)$ should be
bounded at $\pi \downarrow 0$, we must put $C_{01} = C_{10} = 0$ in
(\ref{A33c}). Then, applying the instantaneous-stopping and
smooth-fit conditions from (\ref{instop3}) and (\ref{smfit3}) to the
function in (\ref{A33c}), we get that the equalities:
       \begin{align}
       \label{C01pq0}
       C_{11} \, Q_1(g) - C_{00} \, Q_0(g) = R_0(g) \quad \text{and} \quad C_{11} \, Q_1(h) - C_{00} \, Q_0(h) = R_1(h) \\
       \label{C01pq1}
       C_{11} \, Q_1'(g) - C_{00} \, Q_0'(g) = R_0'(g) \quad \text{and} \quad C_{11} \, Q_1'(h) - C_{00} \, Q_0'(h) = R_1'(h)
       \end{align}
       hold for some $0 < g < h < 1$, where we set:
        \begin{equation}
        \label{R01}
        R_0(\pi) = - a \, \pi + \frac{1-2\pi}{2 \lambda + r}
        \quad \text{and} \quad
        R_1(\pi) = b \, (1 - \pi) + \frac{1-2\pi}{2 \lambda + r}
        \end{equation}
        for all $\pi \in [0, 1]$. Solving the left-hand part of the system in
        (\ref{C01pq0})-(\ref{C01pq1}), we obtain:
\begin{equation}
\label{C01g}
{\hat C}_{00}(g) = \frac{R_0(g) Q_1'(g) - R_0'(g) Q_1(g)}{Q_1(g) Q_0'(g) - Q_1'(g) Q_0(g)}
\quad \text{and} \quad
{\hat C}_{11}(g) = \frac{R_0(g) Q_0'(g) - R_0'(g) Q_0(g)}{Q_1(g) Q_0'(g) - Q_1'(g) Q_0(g)}
\end{equation}
and the solution of the right-hand part there gives:
\begin{equation}
\label{C01h}
{\tilde C}_{00}(h) = \frac{R_1(h) Q_1'(h) - R_1'(h) Q_1(h)}{Q_1(h) Q_0'(h) - Q_1'(h) Q_0(h)}
\quad \text{and} \quad
{\tilde C}_{11}(h) = \frac{R_1(h) Q_0'(h) - R_1'(h) Q_0(h)}{Q_1(h) Q_0'(h) - Q_1'(h) Q_0(h)}
\end{equation}
so that the system in (\ref{C01pq0})-(\ref{C01pq1}) is equivalent
to:
\begin{equation}
\label{BA3}
{\hat C}_{00}(g) = {\tilde C}_{00}(h) \quad \text{and} \quad {\hat C}_{11}(g) = {\tilde C}_{11}(h)
\end{equation}
for $0 < g < h < 1$. It thus follows that the functions:
        \begin{equation}
        \label{U5gh}
        V_0(\pi; g) = {\hat C}_{00}(g) \, Q_0(\pi) + \frac{\lambda + r(1 - \pi)}{r(2\lambda + r)} \quad \text{and} \quad
        V_1(\pi; h) = {\tilde C}_{11}(h) \, Q_1(\pi) + \frac{\lambda + r \pi}{r(2\lambda + r)}
        \end{equation}
        provide a solution of the system in (\ref{VC3})-(\ref{smfit3}) for any $0 < g < h < 1$ fixed.

\vspace{6pt}

        3.2. (Uniqueness in the first formulation.)
        Let us now show that the system in (\ref{BA3}) with (\ref{C01g})-(\ref{C01h}) admits a unique solution $g_*$ and $h_*$.
        For this, using the standard comparison arguments for solutions of the second order ordinary differential
        equations in (\ref{VC3}), we conclude that the resulting curves $\pi \mapsto V_0(\pi; g)$ and $\pi \mapsto V_1(\pi; h)$
        from (\ref{U5gh}) do not intersect each other on the intervals $[g, 1)$ and $(0, h]$, respectively, for different $0 < g < h < 1$ fixed.
        We also observe by virtue of the properties of the functions $Q_{i}(\pi)$, $i = 0, 1$, in (\ref{Q01})
        that $V_0(\pi; g)$ and $V_1(\pi; h)$ are bounded and concave on $[g, 1)$ and $(0, h]$, respectively,
        and such that $V_0'(\pi; g) \to \infty$ as $\pi \downarrow 0$ and $V_1'(\pi; h) \to - \infty$ as $\pi \uparrow 1$.
On the other hand, using the conditions in (\ref{Upi3a}), we obtain
by means of straightforward computations that the inequalities in
(\ref{VD3}) are satisfied whenever $0 < g < {\bar g}$ and ${\bar h}
< h < 1$, where we set:
\begin{equation}
\label{bargh}
{\bar g} = \frac{1+\lambda a}{2+a(2\lambda+r)} \quad \text{and} \quad
{\bar h} = \frac{1 + b(\lambda+r)}{2+b(2\lambda+r)}
\end{equation}
and note that $0 < {\bar g} < 1/2 < {\bar h} < 1$ holds. Hence, we
may conclude that if the conditions:
\begin{equation}
\label{U'bargh}
V_0'({\bar g}+; {\bar g}) < a + V_1'({\bar g}+; {\bar h})
\quad \text{and} \quad
V_1'({\bar h}-; {\bar h}) > - b + V_0'({\bar h}-; {\bar g})
\end{equation}
are satisfied, then the boundaries $g_*$ and $h_*$ belong to the
intervals $(0, {\bar g})$ and $({\bar h}, 1)$, respectively. In
other words, the assumptions in (\ref{U'bargh}) describe the set of
all admissible parameters $a, b > 0$ for which the free-boundary
problem of (\ref{VC3})-(\ref{VD3}) admits a unique solution, so that
the optimal stopping and switching times are given by (\ref{tau13a})
and (\ref{tau13}), respectively.

\vspace{6pt}

       3.3. (Existence in the second formulation.)
       The general solutions of the second order ordinary
       differential equations in (\ref{VC4}) have the form:
       \begin{equation}
       \label{F33c}
       U_i(\pi) = D_{i0} \, G_{i0}(\pi) + D_{i1} \, G_{i1}(\pi)
       + \frac{\lambda}{r(\lambda+r)} + \frac{i \pi}{\lambda+r} + \frac{(1-i)(1-\pi)}{\lambda+r}
       \end{equation}
       where $D_{ij}$ are some arbitrary constants
       and the functions $G_{ij}(\pi)$, $i, j = 0, 1$, are given by:
       \begin{align}
       \label{F33cc}
       G_{00}(\pi) = (1-\pi) \left( \frac{\pi}{1-\pi} \right)^{\gamma_+} 
       \Psi \left( \gamma_+ - 1, \gamma_+ - \gamma_- +1; \frac{2 \lambda \pi}{\rho (1-\pi)} \right) \\
       \label{F33aa}
       G_{01}(\pi) = (1-\pi) \left( \frac{\pi}{1-\pi} \right)^{\gamma_+} 
       \Phi \left( \gamma_+ - 1, \gamma_+ - \gamma_- +1; \frac{2 \lambda \pi}{\rho (1-\pi)} \right)
       \end{align}
       and
       \begin{align}
       \label{F33bb}
       G_{10}(\pi) = \pi \left( \frac{1-\pi}{\pi} \right)^{\gamma_+} 
       \Phi \left( \gamma_+ - 1, \gamma_+ - \gamma_- +1; \frac{2\lambda(1-\pi)}{\rho \pi} \right) \\
       \label{F33dd}
       G_{11}(\pi) = \pi \left( \frac{1-\pi}{\pi} \right)^{\gamma_+} 
       \Psi \left( \gamma_+ - 1, \gamma_+ - \gamma_- +1; \frac{2\lambda(1-\pi)}{\rho \pi} \right)
       \end{align}
       with
       \begin{equation}
       \label{gamma}
       \rho = \left( \frac{\mu_1-\mu_0}{\sigma} \right)^2 \quad \text{and} \quad
       \gamma_{\pm} = \frac{1}{2} + \frac{\lambda}{\rho} \pm \sqrt{ \left( \frac{1}{2} + \frac{\lambda}{\rho} \right)^2 + \frac{2r}{\rho}}
       \end{equation}
       for all $\pi \in (0, 1)$. Here, we denote by:
       \begin{align}
       \label{Phi31}
       \Phi(\alpha, \beta; x) &= 1 + \sum_{k=1}^{\infty}
       \frac{(\alpha)_k}{(\beta)_k} \, \frac{x^k}{k!} \\
       \label{Psi31}
       \Psi(\alpha, \beta; x) &= \frac{\pi}{\sin (\pi \beta)} \left( \frac{\Phi(\alpha, \beta; x)}{\Gamma(1+\alpha-\beta) \Gamma(\beta)}
       - x^{1-\beta} \, \frac{\Phi(1+\alpha-\beta, 2-\beta; x)}{\Gamma(\alpha) \Gamma(2-\beta)} \right)
       \end{align}
       Kummer's confluent hypergeometric functions of the first and second kind,
       respectively, for $\beta \neq 0, -1, -2, \ldots$ and $(\beta)_k = \beta (\beta+1) \cdots (\beta+k-1)$,
       $k \in \NN$, where the series in (\ref{Phi31}) converges under all $x > 0$
       (see, e.g. \cite[Chapter~XIII]{AS} and \cite[Chapter~VI]{BE}),
       and $\Gamma$ denotes Euler's Gamma function. According to the
       results from \cite[Chapter~V, Section~50]{RW}, we can specify the positive
       (strictly) convex functions $G_{i0}(\pi)$, $i = 0, 1$, and $G_{i1}(\pi)$, $i = 0, 1$, as (strictly)
       decreasing and increasing on the interval $(0, 1)$ with 
       singularities at $0$ and $1$, respectively.

       Taking into account the fact that the function $U_0(\pi)$ should be bounded as $\pi \uparrow 1$
       while the function $U_1(\pi)$ should be bounded at $\pi \downarrow 0$, we must put $D_{01} = D_{10} = 0$
       in (\ref{F33c}). Then, applying the instantaneous-stopping and smooth-fit conditions from (\ref{instop4})
       and (\ref{smfit4}) to the function in (\ref{F33c}), we get that the equalities:
        \begin{align}
        \label{C01pq00}
        D_{11} \, G_{11}(p) - D_{00} \, G_{00}(p) = S_0(p) \quad \text{and} \quad D_{11} \, G_{11}(q) - D_{00} \, G_{00}(q) = S_1(q) \\
        \label{C01pq11}
        D_{11} \, G_{11}'(p) - D_{00} \, G_{00}'(p) = S_0'(p) \quad \text{and} \quad D_{11} \, G_{11}'(q) - D_{00} \, G_{00}'(q) = S_1'(q)
        \end{align}
        hold for some $0 < p < q < 1$, where we set:
        \begin{equation}
        \label{D01}
        S_0(\pi) =  - a \, \pi + \frac{1 - 2\pi}{\lambda + r} \quad \text{and} \quad
        S_1(\pi) =  b \, (1 - \pi) + \frac{1 - 2\pi}{\lambda + r}
        \end{equation}
        for all $\pi \in [0, 1]$. Solving the left-hand part of the system in
        (\ref{C01pq00})-(\ref{C01pq11}), we obtain:
\begin{equation}
\label{C0011a}
{\hat D}_{00}(p) = \frac{S_0(p) G_{11}'(p) - S_0'(p) G_{11}(p)}{G_{11}(p) G_{00}'(p) - G_{11}'(p) G_{00}(p)}
\quad \text{and} \quad
{\hat D}_{11}(p) = \frac{S_0(p) G_{00}'(p) - S_0'(p) G_{00}(p)}{G_{11}(p) G_{00}'(p) - G_{11}'(p) G_{00}(p)}
\end{equation}
and the solution of the right-hand part there gives:
\begin{equation}
\label{C0011b}
{\tilde D}_{00}(q) = \frac{S_1(q) G_{11}'(q) - S_1'(q) G_{11}(q)}{G_{11}(q) G_{00}'(q) - G_{11}'(q) G_{00}(q)}
\quad \text{and} \quad
{\tilde D}_{11}(q) = \frac{S_1(q) G_{00}'(q) - S_1'(q) G_{11}(q)}{G_{11}(q) G_{00}'(q) - G_{11}'(q) G_{00}(q)}
\end{equation}
        so that the system in (\ref{C01pq00})-(\ref{C01pq11}) is equivalent to:
\begin{equation}
\label{C0011}
{\hat D}_{00}(p) = {\tilde D}_{00}(q) \quad \text{and} \quad {\hat D}_{11}(p) = {\tilde D}_{11}(q)
\end{equation}
for $0 < p < q < 1$. It thus follows that the functions:
        \begin{equation}
        \label{Vpq}
        U_0(\pi; p) = {\hat D}_{00}(p) \, G_{00}(\pi) + \frac{\lambda+r(1-\pi)}{r(\lambda+r)}
        \quad \text{and} \quad
        U_1(\pi; q) = {\tilde D}_{11}(q) \, G_{11}(\pi) + \frac{\lambda+r\pi}{r(\lambda+r)}
        \end{equation}
        provide a solution of the system in (\ref{VC4})-(\ref{smfit4}) for any $0 < p < q < 1$ fixed.

\vspace{6pt}

3.4. (Uniqueness in the second formulation.) Let us finally follow
the schema of arguments above, to prove that the system of equations
in (\ref{C0011}) with (\ref{C0011a})-(\ref{C0011b}) admits a unique
solution $p_*$ and $q_*$. For this, we use the standard comparison
arguments for solutions of the second order ordinary differential
equations in (\ref{VC4}) to conclude that the curves $\pi \mapsto
U_0(\pi; p)$ and $\pi \mapsto U_1(\pi; q)$ from (\ref{Vpq}) do not
intersect each other on the intervals $[p, 1)$ and $(0, q]$,
respectively, for different $0 < p < q < 1$ fixed. We also observe
by virtue of the properties of the functions $G_{ii}(\pi)$, $i = 0,
1$, in (\ref{F33cc}) and (\ref{F33dd}) that $U_0(\pi; p)$ and
$U_1(\pi; q)$ are bounded and concave on $[p, 1)$ and $(0, q]$,
respectively, and such that $U_0'(\pi; p) \to \infty$ as $\pi
\downarrow 0$ and $U_1'(\pi; q) \to - \infty$ as $\pi \uparrow 1$.
Moreover, using the conditions in (\ref{Upi4a}), we obtain by means
of straightforward computations that the inequalities in (\ref{VD4})
are equivalent to:
\begin{align}
\label{barpq0}
&(2 + a(\lambda+r)) p - \frac{r}{\lambda + r} < - \lambda {\tilde D}_{11}(q) G_{11}'(p) \\ 
\label{barpq1}
&(2 + b(\lambda+r)) (1-q) - \frac{r}{\lambda + r} < \lambda {\hat D}_{00}(p) G_{00}'(q) 
\end{align}
for $0 < p < q < 1$. Note that since the derivative $G_{11}'(\pi)$
is positive and increasing from zero to infinity, while the
derivative $G_{00}'(\pi)$ is negative and increasing from minus
infinity to zero, it is shown by means of standard arguments that
the inequalities in (\ref{barpq0}) and (\ref{barpq1}) hold whenever
$0 < p < {\bar p}$ and ${\bar q} < q < 1$, where the numbers ${\bar
p}$ and ${\bar q}$ are set by:
\begin{equation}
\label{barpq}
{\bar p} = {\hat p} \wedge \frac{r}{(\lambda + r)(2+a(\lambda+r))} < \frac{1}{2} \quad \text{and} \quad
{\bar q} = {\hat q} \vee \frac{\lambda + (\lambda + r)(1 + b(\lambda + r))}{(\lambda + r)(2+b(\lambda+r))} > \frac{1}{2}.
\end{equation}
Here, the couple ${\hat p}$ and ${\hat q}$ is determined as a unique
solution of the corresponding equations instead of the inequalities
in (\ref{barpq0}) and (\ref{barpq1}) whenever it exists, and ${\hat
p} = {\hat q} = 1/2$ otherwise. Hence, we may conclude that if the
conditions:
\begin{equation}
\label{V'barpq}
U_0'({\bar p}+; {\bar p}) < a + U_1'({\bar p}+; {\bar q}) \quad \text{and} \quad
U_1'({\bar q}-; {\bar q}) > - b + U_0'({\bar q}-; {\bar p})
\end{equation}
hold, then the system in (\ref{C0011}) admits a unique solution
$p_*$ and $q_*$ such that $0 < p_* < {\bar p}$ and ${\bar q} < q_* <
1$. Therefore, the assumptions in (\ref{V'barpq}) describe the set
of all admissible parameters $a, b > 0$ for which the free-boundary
problem of (\ref{VC4})-(\ref{VD4}) admits a unique solution, so that
the optimal stopping and switching times are given by
(\ref{tau13a4}) and (\ref{sigma13}), respectively.

  \section{\dot Main results and proofs}

     Taking into account the facts proved above, we are now ready
     to formulate and prove the main assertions of the paper.

\vspace{6pt}

     {\bf Theorem 4.1.}
     {\it Assume that the conditions in (\ref{U'bargh}) are satisfied
     with ${\bar g}$ and ${\bar h}$ defined in (\ref{bargh}).
     Then, in the switching multiple disorder problem of (\ref{V4a})-(\ref{W4a})
     and (\ref{V4b})-(\ref{W4b}) for the process $X$ from (\ref{X4}),
     the Bayesian risk functions $V_i^*(\pi)$, $i = 0, 1$, take the form:
      \begin{align}
      \label{rho5ca}
      V_0^*(\pi) &=
      \begin{cases}
      V_0(\pi; g_*), & \text{if} \; \; g_* < \pi \le 1 \\
      a \, \pi + V_1(\pi; h_*), & \text{if} \; \; 0 \le \pi \le g_* \\
      \end{cases} \\
      \label{rho5cb}
      V_1^*(\pi) &=
      \begin{cases}
      V_1(\pi; h_*), & \text{if} \; \; 0 \le \pi < h_* \\
      b \, (1-\pi) + V_0(\pi; g_*), & \text{if} \; \; h_* \le \pi \le 1
      \end{cases}
      \end{align}
     and the optimal switching times $(\tau^*_{i,n})_{n \in \NN}$,
     $i = 0, 1$, have the structure of (\ref{tau13}). Here, the functions
     $V_0(\pi; g)$ and $V_1(\pi; h)$ are given by (\ref{U5gh}),
     and the optimal stopping boundaries $g_*$ and $h_*$, such that
     $0 < g_* < {\bar g} < 1/2 < {\bar h} < h_* < 1$,
     are uniquely determined by the coupled system of the equations in (\ref{BA3})
     with ${\hat C}_{ii}(g)$ and ${\tilde C}_{ii}(h)$ given by (\ref{C01g})-(\ref{C01h}),
     where the functions $Q_i(\pi)$ and $R_i(\pi)$, $i = 0, 1$,
     are defined in (\ref{Q01}) and (\ref{R01}), respectively.}

     \vspace{6pt}

     {\bf Theorem 4.2.}
     {\it Assume that the conditions in (\ref{V'barpq}) are satisfied
     with ${\bar p}$ and ${\bar q}$ defined by (\ref{barpq}), where
     ${\hat p}$ and ${\hat q}$ is a unique solution of the system of
     equations replacing the inequalities in (\ref{barpq0})-(\ref{barpq1})
     whenever it exists, and ${\hat p} = {\hat g} = 1/2$ otherwise.
     Then, in the switching multiple disorder problem of (\ref{V3a})-(\ref{W3a})
     and (\ref{V3b})-(\ref{W3b}) for the process $X$ from (\ref{X4}),
     the Bayesian risk functions $U^*_i(\pi)$, $i = 0, 1$,
     take the form:
      \begin{align}
      \label{rho3ca}
      U_0^*(\pi) &=
      \begin{cases}
      U_0(\pi; p_*), & \text{if} \; \; p_* < \pi \le 1 \\
      a \, \pi + U_1(\pi; q_*), & \text{if} \; \; 0 \le \pi \le p_*
      \end{cases} \\
      \label{rho3cb}
      U_1^*(\pi) &=
      \begin{cases}
      U_1(\pi; q_*), & \text{if} \; \; 0 \le \pi < q_* \\
      b \, (1-\pi) + U_0(\pi; p_*), & \text{if} \; \; q_* \le \pi \le 1
      \end{cases}
      \end{align}
     and the optimal switching times $(\zeta^*_{i,n})_{n \in \NN}$, $i
     = 0, 1$, have the structure of (\ref{sigma13}). Here, the functions
     $U_0(\pi; p)$ and $U_1(\pi; q)$ are given by (\ref{Vpq}),
     and the optimal stopping boundaries $p_*$ and $q_*$,
     such that $0 < p_* < {\bar p} < 1/2 < {\bar q} < q_* < 1$,
     are uniquely determined by the coupled system
     of the equations in (\ref{C0011}) with ${\hat D}_{ii}(p)$ and ${\tilde D}_{ii}(q)$, $i = 0, 1$,
     given by (\ref{C0011a})-(\ref{C0011b}),
     where the functions $G_{ii}(\pi)$ and $S_i(\pi)$, $i = 0, 1$, are defined in
     (\ref{F33cc})-(\ref{F33dd}) and (\ref{D01}), respectively.}

\vspace{6pt}

        {\bf Proof.} 
        Since the verification of the assertions stated above can be done using
        similar ways of arguments, we present the proof of the second
        one only. Namely, we show that the functions in (\ref{rho3ca})
        and (\ref{rho3cb}) coincide with the value functions in (\ref{V3b})
        and (\ref{W3b}), respectively, and the stopping times $\zeta^*_i$, $i = 0, 1$,
        from (\ref{tau13a4}) and thus the switching times $(\zeta^*_{i,n})_{n \in \NN}$ from (\ref{sigma13})
        are optimal with the boundaries $p_*$ and $q_*$ specified above.
        For this, let us denote by $U_0(\pi)$ and $U_1(\pi)$ the right-hand
        sides of the expressions in (\ref{rho3ca}) and (\ref{rho3cb}), respectively.
        Hence, applying It{\^o}'s formula to $e^{- r t} U_i(\Pi^{(i)}_t)$, $i = 0, 1$, and taking into account
        the smooth-fit conditions in (\ref{smfit4}), we obtain:
        \begin{equation}
        \label{rho4ca}
        e^{- r t} \, U_i(\Pi^{(i)}_t) = U_i(\pi) + \int_0^t e^{- r s} \, (\LL_i U_i - r U_i)(\Pi^{(i)}_s)
        \, I(\Pi^{(i)}_s \neq p_*, \Pi^{(i)}_s \neq q_*) \, ds + M^{(i)}_t
        \end{equation}
        where the processes $M^{(i)} = (M^{(i)}_t)_{t \ge 0}$ defined by:
        \begin{equation}
        \label{M4}
        M^{(i)}_t = \int_0^t e^{- r s} \, U_i'(\Pi^{(i)}_s) \, \frac{\mu_1 - \mu_0}{\sigma} \,
        \Pi^{(i)}_s (1-\Pi^{(i)}_s)
        \, d{\overline B}_s
        \end{equation}
        are continuous square integrable martingales under the probability measure $P_{\pi}$ with respect to
        the filtration $({\cal F}_t)_{t \ge 0}$, for every $i = 0, 1$. The latter fact can easily be observed,
        since the derivatives $U_i'(\pi)$, $i = 0, 1$, are bounded functions.

        Taking into account the assumptions in (\ref{V'barpq}), it is shown by means straightforward
        computations and using the properties of the functions $U_i(\pi)$, $i = 0, 1$,
        that the conditions of (\ref{Cont4c}) and (\ref{VD4}) hold with $0 < p_* < {\bar p}$ and ${\bar q} < q_* < 1$.
        These facts together with the conditions in (\ref{VC4})-(\ref{instop4}) and (\ref{Upi4a}) yield that the inequalities
        $(\LL_0 U_0 - r U_0)(\pi) \ge - (1-\pi)$ and $(\LL_1 U_1 - r U_1)(\pi) \ge - \pi$ hold for all $\pi \in [0, 1]$
        such that $\pi \neq p_*$ and $\pi \neq q_*$, as well as $U_0(\pi) \le a \pi + U_1(\pi)$ and
        $U_1(\pi) \le b (1-\pi) + U_0(\pi)$ are satisfied for all $\pi \in [0, 1]$.
        It also follows from the regularity of the diffusion processes
        $\Pi^{(i)}$, $i = 0, 1$, solving the stochastic differential
        equations in (\ref{pi4ca}) and (\ref{pi4cb}), that the
        indicator which appears in the formula (\ref{rho4ca}) can be ignored.
%
        We therefore obtain from the expression in (\ref{rho4ca}) 
        that the inequalities:
        \begin{align}
        \label{rho4ea}
        &a \, e^{-r \zeta_0} \, \Pi^{(0)}_{\zeta_0} + \int_0^{\zeta_0} e^{- r s} \, (1-\Pi^{(0)}_s) \, ds + e^{-r \zeta_0} \, U_1(\Pi^{(0)}_{\zeta_0}) \\
        \notag
        &\ge e^{-r \zeta_0} \, U_0(\Pi^{(0)}_{\zeta_0}) + \int_0^{\zeta_0} e^{- r s} \, (1-\Pi^{(0)}_s) \, ds \ge U_0(\pi) + M^{(0)}_{\zeta_0}
        \end{align}
        \vspace{-13pt}
        \begin{align}
        \label{rho4eb}
        &b \, e^{- r \zeta_1} \, (1-\Pi^{(1)}_{\zeta_1}) + \int_0^{\zeta_1} e^{- r s} \, \Pi^{(1)}_s \, ds + e^{- r \zeta_1} \, U_0(\Pi^{(1)}_{\zeta_1}) \\
        \notag
        &\ge e^{- r \zeta_1} \, U_1(\Pi^{(1)}_{\zeta_1}) + \int_0^{\zeta_1} e^{- r s} \, \Pi^{(1)}_s \, ds \ge U_1(\pi) + M^{(1)}_{\zeta_1}
        \end{align}
        hold for any stopping times $\zeta_i$ of the processes $\Pi^{(i)}$, $i = 0, 1$, respectively.

        For every $i = 0, 1$, let $(\kappa_{i,n})_{n \in \NN}$
        be an arbitrary localizing sequence of stopping times
        for the processes $M^{(i)}$. 
        Then, taking the expectations with respect to the probability measure $P_{\pi}$
        in (\ref{rho4ea})-(\ref{rho4eb}), by means of the optional sampling theorem
        (see, e.g. \cite[Theorem~3.6]{LS} or \cite[Chapter~I, Theorem~3.22]{KS}), we get:
        \begin{align}
        \label{rho4ec}
        &E_{\pi} \left[ a \, e^{-r (\zeta_0 \wedge \kappa_{0,n})} \, \Pi^{(0)}_{\zeta_0 \wedge \kappa_{0,n}} +
        \int_0^{\zeta_0 \wedge \kappa_{0,n}} e^{- r s} \, (1-\Pi^{(0)}_s) \, ds +
        e^{-r (\zeta_0 \wedge \kappa_{0,n})} \, U_1(\Pi^{(0)}_{\zeta_0 \wedge \kappa_{0,n}}) \right] \\
        \notag
        &\ge E_{\pi} \left[ e^{-r (\zeta_0 \wedge \kappa_{0,n})} \, U_0(\Pi^{(0)}_{\zeta_0 \wedge \kappa_{0,n}}) +
        \int_0^{\zeta_0 \wedge \kappa_{0,n}} e^{- r s} \, (1-\Pi^{(0)}_s) \, ds \right]
        \ge U_0(\pi) + E_{\pi} \big[ M^{(0)}_{\zeta_0 \wedge \kappa_{0,n}} \big] = U_0(\pi)
        \end{align}
        \vspace{-13pt}
        \begin{align}
        \label{rho4ed}
        &E_{\pi} \left[ b \, e^{- r (\zeta_1 \wedge \kappa_{1,n})} \, (1-\Pi^{(1)}_{\zeta_1 \wedge \kappa_{1,n}}) +
        \int_0^{\zeta_1 \wedge \kappa_{1,n}} e^{- r s} \, \Pi^{(1)}_s \, ds + e^{-r (\zeta_1 \wedge \kappa_{1,n})} \,
        U_0(\Pi^{(1)}_{\zeta_1 \wedge \kappa_{1,n}}) \right] \\
        \notag
        &\ge E_{\pi} \left[ e^{- r (\zeta_1 \wedge \kappa_{1,n})} \, U_1(\Pi^{(1)}_{\zeta_1 \wedge \kappa_{1,n}}) +
        \int_0^{\zeta_1 \wedge \kappa_{1,n}} e^{- r s} \, \Pi^{(1)}_s \, ds \right]
        \ge U_1(\pi) + E_{\pi} \big[ M^{(1)}_{\zeta_1 \wedge \kappa_{1,n}} \big] = U_1(\pi)
        \end{align}
        for all $\pi \in [0, 1]$.
        Thus, letting $n$ go to infinity and using Fatou's lemma,
        we obtain that the inequalities:
        \begin{align}
        \label{rho4ee}
        E_{\pi} \left[ a \, e^{- r \zeta_0} \, \Pi^{(0)}_{\zeta_0} + \int_0^{\zeta_0} e^{- r s} \, (1-\Pi^{(0)}_s) \, ds +
        e^{- r \zeta_0} \, U_1(\Pi^{(0)}_{\zeta_0}) \right] &\ge U_0(\pi) \\
        \label{rho4ef}
        E_{\pi} \left[ b \, e^{- r \zeta_1} \, (1-\Pi^{(1)}_{\zeta_1}) + \int_0^{\zeta_1}
        e^{- r s} \, \Pi^{(1)}_s \, ds + e^{- r \zeta_1} \, U_0(\Pi^{(1)}_{\zeta_1}) \right] &\ge U_1(\pi)
        \end{align}
        are satisfied for any stopping times $\zeta_i$, $i = 0, 1$, and all $\pi \in [0, 1]$.
        By virtue of the structure of the stopping times in (\ref{tau13a4}), it is
        readily seen that the equalities in (\ref{rho4ee}) and (\ref{rho4ef}) hold with $\zeta^*_i$
        instead of $\zeta_i$, $i = 0, 1$, when either $\pi \le p_*$ or $\pi \ge q_*$, respectively.

       It remains to show that the equalities are attained in (\ref{rho4ee}) and (\ref{rho4ef})
       when $\zeta^*_i$ replaces $\zeta_i$, $i = 0, 1$, for $p_* < \pi < q_*$.
       By virtue of the fact that the functions $U_i(\pi)$, $i = 0, 1$,
       with the boundaries $p_*$ and $q_*$ satisfy the conditions
       in (\ref{VC4}) and (\ref{instop4}), it follows from the
       expression in (\ref{rho4ca}) and the structure of the
       stopping times in (\ref{tau13a4}) that the equalities:
        \begin{align}
        \label{rho4ga}
        &e^{- r (\zeta_0^* \wedge \kappa_{0,n})} \, U_0(\Pi^{(0)}_{\zeta_0^* \wedge \kappa_{0,n}})
        + \int_0^{\zeta_0^* \wedge \kappa_{0,n}} e^{- r s} \, (1-\Pi^{(0)}_s) \, ds
        = U_0(\pi) + M^{(0)}_{\zeta_0^* \wedge \kappa_{0,n}} \\
        \label{rho4gb}
        &e^{- r (\zeta_1^* \wedge \kappa_{1,n})} \, U_1(\Pi^{(1)}_{\zeta_1^* \wedge \kappa_{1,n}}) +
        \int_0^{\zeta_1^* \wedge \kappa_{1,n}} e^{- r s} \, \Pi^{(1)}_s \, ds
        = U_1(\pi) + M^{(1)}_{\zeta_1^* \wedge \kappa_{1,n}}
        \end{align}
        are satisfied for all $\pi \in [0, 1]$. Observe that the integrals here are finite ($P_{\pi}$-a.s.)
        as well as the processes $(M^{(i)}_{\zeta_i^* \wedge t})_{t \ge 0}$, $i = 0, 1$, are uniformly integrable martingales.
        Therefore, taking the expectations in (\ref{rho4ga}) and (\ref{rho4gb}) and letting $n$ go to infinity,
        we can apply the Lebesgue dominated convergence theorem to obtain the equalities:
        \begin{align}
        \label{rho4ia}
        &E_{\pi} \left[ a \, e^{- r \zeta_0^*} \, \Pi^{(0)}_{\zeta_0^*} +
        \int_0^{\zeta_0^*} e^{- r s} \, (1-\Pi^{(0)}_s) \, ds
        + e^{- r \zeta_0^*} \, U_1(\Pi^{(0)}_{\zeta_0^*}) \right] = U_0(\pi) \\
        \label{rho4ib}
        &E_{\pi} \left[ b \, e^{- r \zeta_1^*} \, (1-\Pi^{(1)}_{\zeta_1^*}) +
        \int_0^{\zeta_1^*} e^{- r s} \, \Pi^{(1)}_s \, ds
        + e^{- r \zeta_1^*} \, U_0(\Pi^{(1)}_{\zeta_1^*}) \right] = U_1(\pi)
        \end{align}
        for all $\pi \in [0, 1]$. The latter, together with the inequalities in
        (\ref{rho4ee}) and (\ref{rho4ef}), directly imply the desired assertion. $\square$





        \vspace{6pt}

        {\bf Remark 4.3.}
        The results formulated above show that the following sequential procedure is optimal.
        Being based on the observations of $X$, we construct the posterior probability process $\Pi$
        and stop the latter for the first time as soon as it exits either the region $(g_*, h_*)$ or
        $(p_*, q_*)$, appropriately, and then conclude that the process $\Theta$ has switched
        either from $0$ to $1$ or from $1$ to $0$, respectively.
        Then, we continue to observe the process $\Pi$ which is currently located either in the regions
        $[0, g_*]$ and $[h_*, 1]$ or in $[0, p_*]$ and $[q_*, 1]$, and stop the observations as soon as it
        comes to the opposite region. We may thus conclude that $\Theta$ should have switched either
        from $0$ to $1$ or from $1$ to $0$, respectively, and continue the procedure from the beginning.


\vspace{6pt}

  {\bf Remark 4.4.}
  Taking into account the results obtained above, we may also conclude that the appropriate {\it minimal} Bayesian risk functions
  take the form:
  \begin{equation}
  \label{J3}
  V^*(\pi) = \min \{ V^*_0(\pi), V^*_1(\pi) \} \quad \text{and} \quad
  U^*(\pi) = \min \{ U^*_0(\pi), U^*_1(\pi) \}
  \end{equation}
  for $\pi \in [0, 1]$, where the functions $V^*_i(\pi)$ and $U^*_i(\pi)$, $i = 0, 1$,
  are defined in (\ref{V4a})-(\ref{W4a}) and (\ref{V3b})-(\ref{W3b}), respectively.
  It is also seen that if either $V^*(\pi) = V^*_i(\pi)$ or $U^*(\pi) =
  U^*_i(\pi)$ holds for any $\pi \in [0, 1]$ fixed, then the
  sequences $(\tau^*_{i, n})_{n \in \NN}$ or $(\zeta^*_{i, n})_{n \in \NN}$ given by
  (\ref{tau13}) and (\ref{sigma13}) are optimal in (\ref{J3}), appropriately.


\vspace{12pt}

{\bf Acknowledgments.}
The author is grateful to Savas Dayanik for many useful comments and
to Xin Guo for her helpful suggestions. This research was partially
supported by Deutsche Forschungsgemeinschaft through the SFB 649
Economic Risk.


    \begin{center}
    
    \end{center}

\end{document}